\documentclass[aps,prl,twocolumn,superscriptaddress,groupedaddress,amsmath,amssymb,
 aps]{revtex4}  
\usepackage{graphicx}  
\usepackage{dcolumn}   
\usepackage{bm}        
\usepackage{amssymb}   
\usepackage{verbatim}

\newcommand{\kk}{\mathbf{k}}

\begin{document}

\widetext


\title{Amplitude mode oscillations in pump-probe photoemission spectra from a \textit{d}-wave superconductor}
\author{B. Nosarzewski}
 \affiliation{Department of Physics, Stanford University, Stanford, CA 94305, USA }
 \affiliation{Stanford Institude for Materials and Energy Sciences (SIMES), SLAC National Accelerator Laboratory, Menlo Park, CA 94025, USA}

\author{B. Moritz}%
 \affiliation{Stanford Institude for Materials and Energy Sciences (SIMES), SLAC National Accelerator Laboratory, Menlo Park, CA 94025, USA}
 \affiliation{Department of Physics and Astrophysics, University of North Dakota, Grand Forks, ND 58202}

\author{J. K. Freericks}
\affiliation{Department of Physics, Georgetown University, Washington, DC 20057, USA}

\author{A. F. Kemper}%
 \affiliation{Department of Physics, North Carolina State University, Raleigh, NC 27695, USA}

\author{T. P. Devereaux}%
 \affiliation{Stanford Institude for Materials and Energy Sciences (SIMES), SLAC National Accelerator Laboratory, Menlo Park, CA 94025, USA}
 \affiliation{Geballe Laboratory for Advanced Materials, Stanford University, Stanford, CA 94305}

\date{\today}

\begin{abstract}
Recent developments in the techniques of ultrafast pump-probe photoemission have made possible the search for collective modes in strongly correlated systems out of equilibrium. Including inelastic scattering processes and a retarded interaction, we simulate time- and angle- resolved photoemission spectroscopy (trARPES) to study the amplitude mode of a \textit{d}-wave superconductor, a collective mode excited through the nonlinear light-matter coupling to the pump pulse. We find that the amplitude mode oscillations of the \textit{d}-wave order parameter occur in phase at a single frequency that is twice the quasi-steady-state maximum gap size after pumping. We comment on the necessary conditions for detecting the amplitude mode in trARPES experiments.

\end{abstract}

\maketitle

The amplitude mode of the superconducting order parameter, also known as the Higgs mode, is fundamental to superconductivity and arises because of the broken gauge symmetry of the superconducting state. Observing this mode is interesting from the perspective of understanding the collective behavior of a macroscopic quantum state out of equilibrium and has been the subject of several experimental studies performed on \textit{s}-wave superconductors using Raman and THz pump-probe spectroscopy \cite{Raman, Raman2, Shimano}. However, these experimental techniques are most likely not as well-suited as the quickly advancing technique of time- and angle- resolved photoemission spectroscopy (trARPES) for studying the Higgs mode in materials such as the high-$T_c$ cuprate superconductors which have a superconducting order parameter with \textit{d}-wave symmetry. We demonstrate that future trARPES experiments may be an ideal candidate to address the question of whether the Higgs mode of a \textit{d}-wave superconductor appears as a single amplitude mode associated with the value of the superconducting gap maximum or as a spectrum of modes arising from the nodal nature of the superconducting order. 


Since the Higgs mode is a scalar boson without charge or spin, it does not couple linearly to electromagnetic fields and is difficult to observe via the standard experimental probes of the equilibrium state \cite{Varma, Podolsky}. Traditionally the Higgs mode has been detected indirectly through Raman spectroscopy which relies on the interpretation that the observed $2\Delta$ excitations borrow Raman activity from the coexisting charge density wave via electron-phonon coupling \cite{Raman, Raman2}. As an alternative to probing the equilibrium state, recent advancements in time-domain spectroscopies make possible the direct detection of amplitude modes by driving systems out of equilibrium \cite{CDW}. In a pump-probe experiment, an ultrashort pump pulse excites the system to a nonequilibrium state for which the original magnitude of the order parameter in the equilibrium state is no longer a minimum of the free energy. Because the order is partially melted by a pump pulse, the amplitude mode appears as the oscillation of the order parameter about a new, smaller value due to the decrease in quasiparticles involved in ordering \cite{QQ1}. \par

The first time-domain experiment to successfully detect the Higgs mode was a terahertz pump-probe measurement of the optical conductivity in the \textit{s}-wave superconductor $\text{Nb}_{\text{0.8}}\text{Ti}_{\text{0.2}}\text{N}$ \cite{Shimano}. However, this technique does not straightforwardly provide information about the momentum dependence of the Higgs mode in a \textit{d}-wave superconductor. In contrast, the emerging pump-probe technique of time- and angle- resolved photoemission spectroscopy (trARPES) is an ideal candidate for studying the Higgs mode in \textit{d}-wave superconductors since it is a nonequilibrium technique with the time, energy, and momentum resolution required to directly probe the Higgs mode by observing the behavior of the superconducting gap size on a femtosecond timescale across the Brillouin zone \cite{Lex}. Recently, trARPES has successfully been used to study unoccupied bandstructure, relaxation dynamics, and collective modes in various materials, providing new information beyond the reach of equilibrium spectroscopies \cite{CDW,ShenInequivalence,Lanzara1,Lanzara2,Rameau, Rameau2,Bovensiepen,Perfetti,Shin,Cavalleri,Beyer,Madan,Mansart,Torchinsky,Chia}. For instance, trARPES was used to directly probe the single-particle spectral function and observe the amplitude mode corresponding to oscillations of the charge-density wave order parameter \cite{CDW}.


Our previous work demonstrates that time- and angle- resolved photoemission spectroscopy in principle provides a direct way to detect the Higgs mode \cite{Lex}. We extend upon this work here, building upon the same formalism which simulates the pump-probe process by self-consistently solving the Nambu-Gor'kov equations within the Migdal-Eliashberg approximation. Such a treatment of the interactions and pump process goes beyond the typical theoretical methods used to study the Higgs mode. Previous theoretical work often relies on the simple framework of Bardeen-Cooper-Schrieffer (BCS) theory which neglects inelastic scattering processes and is often limited to performing a quantum quench of the pairing interaction which neglects important dynamical processes present in real materials such as the melting of superconducting order by the pump pulse \cite{VarmaDwave,Capone,QQ1,QQ2,Aoki}. In contrast, our calculation includes inelastic scattering processes (which are important for the dynamics out of equilibrium) and a frequency dependent pairing interaction. Within this framework, we investigate the effect of \textit{d}-wave pair symmetry on the characteristics of the Higgs mode. Our calculation also naturally captures the dynamic process of melting of the superconducting state by the pump pulse and subsequent relaxation due to electron-boson scattering. \par

We solve the time-dependent equations of motion for the Holstein model with a momentum-dependent electron-boson coupling \cite{Holstein}

\begin{equation}
\begin{split}
\mathcal{H} = \sum_\textbf{k} \epsilon_\textbf{k}c^\dagger_\textbf{k}&c^{\phantom{}}_\textbf{k} + \Omega \sum_\textbf{k} b^\dagger_\textbf{k}b^{\phantom{}}_\textbf{k}\\&- \frac{1}{\sqrt{N}}\sum_\textbf{k,q}g(\textbf{k},\textbf{q})c^\dagger_{\textbf{k}-\textbf{q}}c^{\phantom{}}_\textbf{k}(b^\dagger_\textbf{q} + b^{\phantom{}}_{-\textbf{q}}).
\end{split}
\end{equation}
The trARPES spectrum is obtained from the double-time lesser Green's function on the Kadanoff-Baym-Keldysh contour and a Gaussian probe pulse of width $\sigma_p$=16 fs, the gauge-invariant trARPES intensity at time $t_0$ is given by \cite{Freericks}: 

\begin{align}
I(\kk,\omega,t_0) =\mathrm{Im} \int dt\, dt'\, 
p(t,t',t_0)
e^{i\omega(t-t')} G_{\tilde\kk(t,t')}^<(t,t') 
\label{eq:trarpes}
\end{align}
where $p(t,t',t_0)$ is a two-dimensional normalized Gaussian with a width $\sigma_p$ centered at $(t,t')=(t_0,t_0)$ and $G_{\tilde\kk(t,t')}^<(t,t')$ is the lesser Green's function. To calculate the double-time Green's functions on the contour, we self-consistently solve the Nambu-Gor'kov equations of motion. We use units where $c=\hbar=e=1$. The coupling to the field is treated semi-classically and included to all orders via the Peierls’ substitution $\textbf{k}(t)=\textbf{k}-\textbf{A}(t)$ where \textbf{A}(t) is the time varying vector potential in the Hamiltonian gauge. We ensure that the single-particle ARPES spectra are gauge invariant by performing the constructive transformation described by Ref.\ \cite{Bertoncini} which gauge shifts the momentum variable of the Green's function. For a more complete description of the equations of motion and gauge shifting procedure see the Supplemental Material \cite{supplement}. The field of the pump pulse in all simulations below is applied along the diagonal direction of the Brillouin zone and takes the form of a sinusoidal oscillation (energy of 1.5 eV) with a Gaussian envelope (FWHM of 9.3 fs).
\par

Superconductivity in our model is mediated through a generic bosonic mode which is included in the electron self-energy at the self-consistent Born level. This self-energy is given by
\begin{equation}
 \Sigma_\textbf{k}^c(t,t')=\frac{i}{N_k}\sum_{\textbf{k}'} |g(\textbf{k},\textbf{k}')|^2 \tau_3 G_{\textbf{k}'}^c(t,t') \tau_3 D_0^c(t,t'), 
\end{equation}
where $D_0^c$ is the bare propagator for a bosonic mode with frequency $\Omega$, $N_k$ is the number of momenta, $\tau_3$ is the $z$-direction Pauli matrix in Nambu space, and the superscript $c$ indicates contour-ordering on the Kadanoff-Baym-Keldysh contour. We consider $|g(\textbf{k},\textbf{k}')|^2 = g_s + g_dd_\textbf{k}d_{\textbf{k}'}$ where $g_s$ and $g_d$ are constants which set the electron-boson coupling strength, $d_\textbf{k}=\frac{1}{2}[\cos(k_x)-\cos(k_y)]$ is a momentum-dependent form-factor with \textit{d}-wave symmetry, and $\textbf{k}'=\textbf{k}-\textbf{q}$. We work under the desired ansatz that the superconducting state has purely \textit{d}-wave symmetry at all times and do not consider the possibility of changes to the symmetry of the order parameter upon pumping.  We verify that the electron-boson interaction strength remains constant at all times (assuming the boson is unrenormalized and behaves as an infinite heat bath) by checking that the zeroth-moment of the retarded self-energy given by $\Sigma^R(t,t)$ (which is proportional to the square of the coupling strength) is constant \cite{LexPRB90}. Therefore changes in the trARPES spectrum are not a result of changing the electron-boson coupling as in a quantum quench, but instead a consequence of redistribution of spectral weight by the pump and transient effective electron-boson interactions which are determined self-consistently. 

\begin{figure}
\includegraphics[width=\columnwidth]{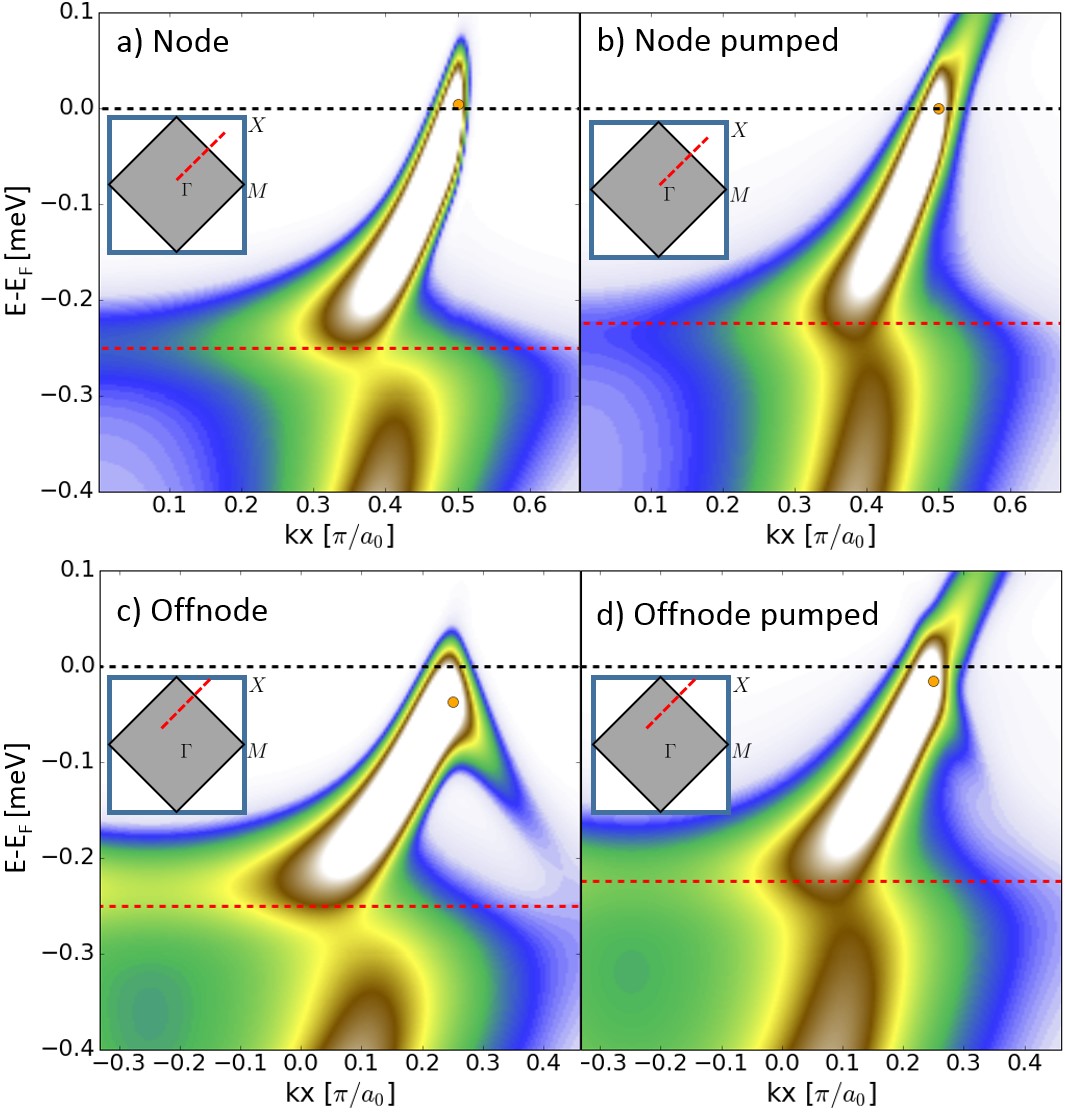}
\caption{\label{fig1} \textbf{trARPES spectra.} In all panels, the orange marker indicates the peak position of the EDC at $k_F$ which corresponds to the superconducting gap size, and the red dotted line indicates the energy of the bosonic mode plus the antinodal gap size which corresponds to the gap-shifted position of the kink in the bandstructure. a) Nodal ARPES spectrum in equilibrium with superconductivity.   b) Nodal spectrum 25 fs after pump arrives. c) Off-nodal ARPES spectrum in equilibrium with superconductivity. d) Pumped off-nodal spectrum shows shift in the kink and a partial melting of the superconducting gap.}
\end{figure}

The equations of motion are solved by performing a massively parallel computation following the methods in Ref.\ \cite{Leeuwen}. For ease of simulation, we take a tight-binding model on a square lattice at half-filling with a nearest-neighbor hopping of $V_{nn} = 0.25$ eV. We take a mode energy of $\Omega = 0.2$ eV, a temperature of $T \simeq 80 K$, and a coupling strength of $g_s=g_d=\sqrt{0.12}$ eV which results in a dimensionless coupling of $\lambda_s \equiv -\partial \text{Re} \Sigma^R(\omega)/\partial{\omega} |_{\omega \rightarrow 0} = 0.67$. The temperature is well below the transition temperature of $T_c \simeq 240$ K. These parameters are not meant to represent a specific material with a realistic set of parameters, but our results are nevertheless illustrative of pump-probe ARPES spectroscopy on a material with a \textit{d}-wave superconducting order parameter. The large superconducting gap size resulting from these parameters increases the Higgs frequency which reduces the required time to simulate a Keldysh contour long enough to clearly identify the Higgs mode. 
 
In Fig.\ 1(a), we show the ARPES spectrum in equilibrium near the Fermi level along a nodal cut (diagonal cut) which shows no gap at the Fermi level, as expected, and a kink in the bandstructure at the bosonic mode energy (200 meV) gap-shifted by the maximum of the superconducting gap size (51 meV) \cite{Sandvik, DevereauxReview}. The superconducting gap size can be obtained from the equilibrium self-energies, shown in the Supplemental Material \cite{supplement}, or directly from the peak position of an antinodal energy-density curve (EDC) at $k_F$. For an off-nodal cut we take a cut parallel to the zone diagonal and halfway between the node and antinode as marked by the dotted red line in the insets of Fig. 1(c,d). The spectrum for the off-nodal cut is shown in Fig.\ \ref{fig1}(c) and shows a gap at the Fermi level and a clear bend-back of the band due to particle-hole mixing. To determine the superconducting gap size, we find the peak position of the energy density curves (EDCs) at $k=k_F$ by fitting to a Voigt profile. To determine the kink position, we use the energy of the inflection point in the Engelsberg-Schrieffer peak-dip-hump structure of the EDCs at $k=k_F$ \cite{Sandvik}. The gap position and kink positions are indicated with the colored markers and dotted lines in Fig.\ 1. We track these features as a function of time in the trARPES spectra. Figure \ref{fig1}(b,d) show the spectra 25 fs after the center of the pump pulse arrives. Spectral weight is redistributed above the Fermi level and the superconducting gap partially melts, also shifting the kink position to a higher binding energy. \par 

\begin{figure}
\includegraphics[width=\columnwidth]{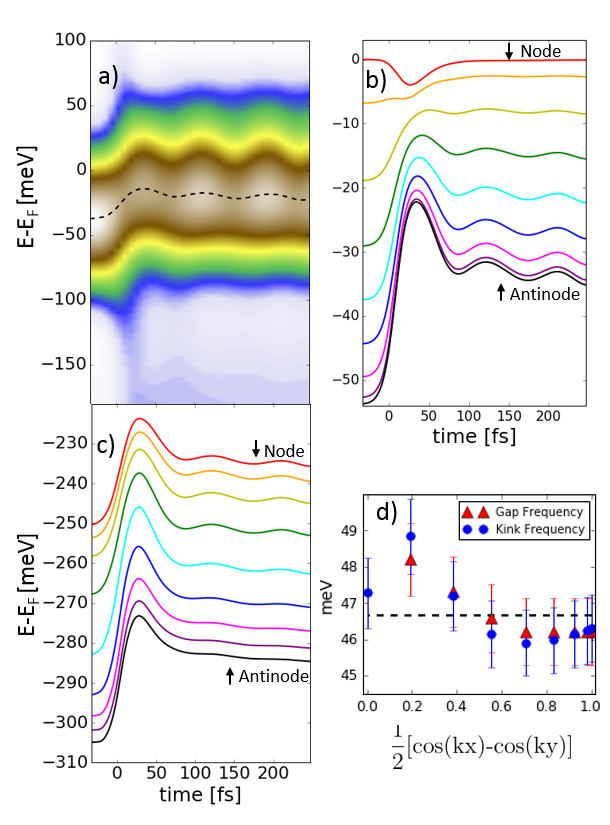}
\caption{\label{fig2} \textbf{Gap and kink dynamics.} a) Tracking the peak position of the EDC at $k_F$ given by the orange marker in Fig. 1 (i.e. superconducting gap size) for an off-nodal cut which is halfway between the node and antinode as indicated by the dotted red line in the inset of Fig 1(c,d). b) Tracking the EDC peak position for multiple cuts along the Fermi surface. c) Tracking the kink position given by the horizontal red dotted line in Fig. 1 based on the inflection point in the peak-dip-hump structure of EDC’s at $k_F$ for multiple cuts. Curves for the kink position are offset for clarity. The pump field is $E_{max}=1.2$ $\text{V}/a_0$ for all plots. The time is measured relative to the center of the pump pulse (which reaches the sample at time $t=0$ fs). d) The frequency of the oscillations (of the gaps and the kinks) occurs at a single frequency given by the average value of 47 meV, shown as the dotted black line. }
\end{figure}

After the pump pulse, clear oscillations occur both in the photoemission spectrum for the range of energies between the gap edge and in the kink position for various momenta along the Fermi surface as shown in Fig. \ref{fig2}. These oscillations are the signatures of the amplitude (or Higgs) mode and result from the oscillation of the magnitude of the superconducting order parameter. As shown in the Supplemental Material \cite{supplement}, oscillations with the same frequency also occur in the anomalous density and the electronic self-energies. Previous work has considered how the Higgs mode is affected by the continuum of single-particle excitations which exhibit a square-root singularity at $2\Delta$ \cite{Benfatto}. We note that the presence of oscillations at the kink position (roughly $4\Delta$ away from the gap-edge) implies that it is not possible to attribute these oscillations to $2\Delta$ quasiparticle excitations. Oscillations of the kink position are expected because the kink position is set by the size of the superconducting gap plus the antinodal gap size. Furthermore, the normal state spectra after pumping return monotonically to equilibrium, indicating that the superconducting order is responsible for the oscillations \cite{Lex}. The oscillations of the gap become weaker and disappear towards the nodal point since the gap size shrinks to zero at the node. However, the value of the EDC maximum in Fig.\ \ref{fig2}(b) is not identically zero at the node because of broadening of the single-particle spectrum due to finite energy resolution.  

The frequencies of the gap and kink oscillations are extracted by fitting to a decaying exponential plus a damped oscillation of the form:

\begin{equation}
\label{eq:fitfunc}
 A e^{-t/\tau} + B \sin(\omega t+\phi)/t^p + D.
\end{equation}
When used to fit the gap position, the parameter D gives $\Delta_\infty$, the quasi-steady-state value of the superconducting gap after the pump pulse (such that the Higgs frequency satisfies $\omega=2\Delta_\infty$). When used to fit the kink position, the parameter D is given by $\Omega+\Delta_\infty$. Our fits of the gap position do not extend all the way to the nodal point because as the gap value becomes smaller the oscillations decrease in amplitude and become more difficult to fit. However, the kink oscillations can still be fit at the node. From the combined analysis of both gap and kink position, we find that the Higgs oscillations occur at a single frequency and in phase across all momentum points within our frequency and energy resolution, as shown in Fig.\ \ref{fig2}d. 
 
\begin{figure}
\includegraphics[scale=0.60]{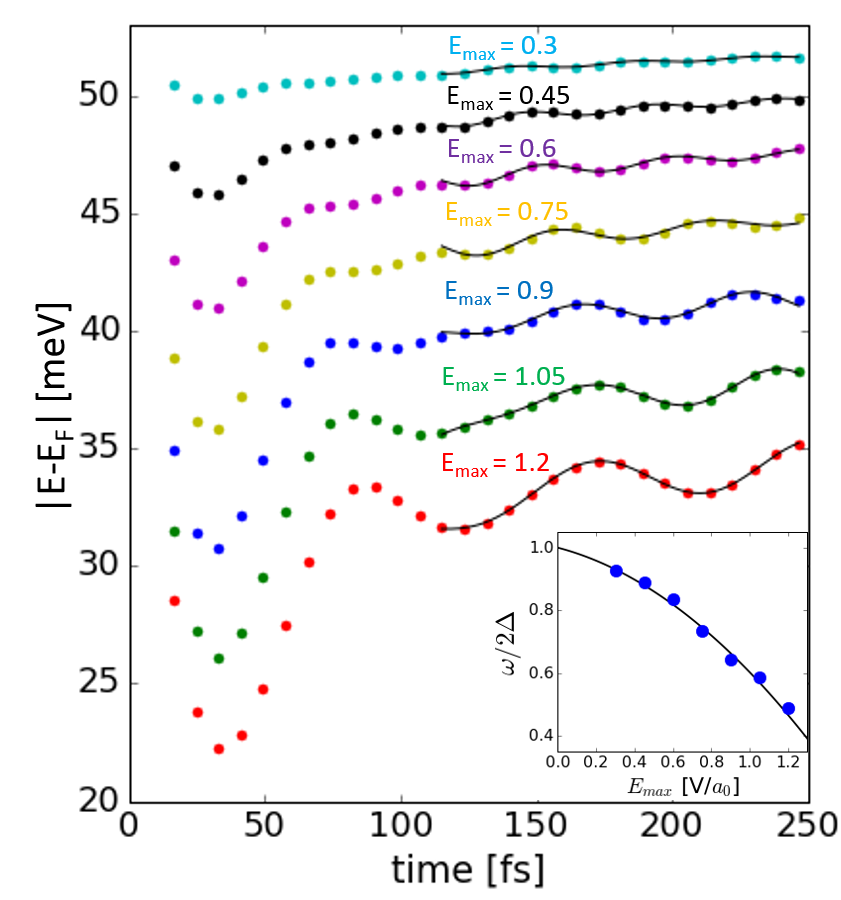}
\caption{\label{fig3} \textbf{Gap dynamics vs. fluence.} The superconducting gap size (determined by the magnitude of the antinodal EDC peak position) as a function of time for different pump fluences (maximum electric field in $V/a_0$). Solid lines show the fits. Inset: in the zero fluence limit, the Higgs frequency extrapolates to twice the maximum gap size in equilibrium. The solid line is a quadratic polynomial fit. }
\end{figure}

In Fig.\ \ref{fig3}, we again use the functional form in Eq.\ (\ref{eq:fitfunc}) to fit the EDC peak position at the antinode for different pump fluences. We note that our simulation requires relatively high field strengths to reach the same regimes that would be reached experimentally in real systems because our model does not consider quantum fluctuations and we choose parameters which result in robust superconducting order with a high $T_c$. We observe that the Higgs oscillation frequency decreases with increasing fluence because the superconducting gap size is suppressed more for stronger pumping \cite{Lex}. The frequency of the Higgs oscillation follows twice the quasi-steady state value of the antinodal (maximum) gap size after pumping ($\Delta_\infty$), and the Higgs frequency extrapolates to twice the value of the antinodal superconducting gap size in equilibrium as shown in the inset in Fig.\ \ref{fig3}. In other words, for a \textit{d}-wave superconductor, the Higgs mode is a ${2\Delta_\infty}$ oscillation with ${\Delta_\infty}$ given by the maximum gap size after pumping. 
\par

The Higgs mode has yet to be detected in trARPES experiments which have up to this point mainly focused on relaxation dynamics of quasiparticles and other collective modes \cite{CDW,ShenInequivalence,LanzaraGS,Lanzara1,Lanzara2,Rameau,Rameau2,Bovensiepen,Perfetti,Shin,Cavalleri,ShuolongPhonon}. In order to satisfy the experimental conditions necessary for observing the Higgs mode, the fluence of the pump pulse must be tuned and both the pump and the probe pulse must be sufficiently fast. Naturally, if the pump is too weak, the amplitude of the Higgs mode will be small. If the pump is too strong, the condensate is fully depleted and Cooper pairs are not available to participate in the collective mode. From our simulations, approximately depleting half of the condensate results in the strongest Higgs oscillations. Recent experiments on $\text{Bi}_2\text{Sr}_2\text{CaCu}_2\text{O}_{\text{8}+\delta}$ (Bi2212) with $T_c=91$ K indicate that a pump fluence of approximately 8 $\mu \text{J}/\text{cm}^2$ suppresses the superconducting gap to half of its equilibrium value \cite{LanzaraGS}. To determine the required width of the pump and probe pulses, we must consider the timescale of the Higgs mode which is set by  $\tau = \textit{h}/2\Delta$. For the pump pulse to nonadiabatically excite the condensate, the width of the pump pulse must be less than $\tau$. In addition, for the probe pulse to resolve the oscillations, the width of the probe pulse must also be less than $\tau$. For a typical cuprate superconductor such as Bi2212 around optimal doping, the equilibrium superconducting gap size is of the order of 40 meV \cite{Inna}. If the gap is suppressed by 50 percent after pumping, the timescale of the Higgs mode will be on the order of $\tau=200$ fs. It is promising that several experimental groups have achieved sub-100fs time resolution \cite{ShenInequivalence, Bovensiepen, Wolf}. A separate factor which could potentially prevent the detection of the Higgs mode in some systems is the presence of inter-band transitions which could destroy the coherent nature of the collective mode. How the Higgs mode appears in multi-band systems with dipole transitions should be clarified in the future.

The stage is set to take advantage of pump-probe techniques such as trARPES not only to detect the Higgs mode, but also to study the rich assortment of collective modes which have been predicted in the unconventional superconductors. Examples include the Bardasis-Schrieffer modes in systems with competition for superconducting ground states with different pairing symmetries \cite{DevereauxBS}, Leggett modes in multi-band superconductors \cite{LeggettSchnyder,LeggettHuang}, multiple Higgs modes in channels corresponding to different irreducible representations of the lattice \cite{VarmaDwave}, and the various collective modes arising in gauge theories \cite{Lee}. Our work serves as a starting point for studying these modes within a framework that includes inelastic scattering and retarded interactions, ingredients which are needed to accurately simulate the amplitude mode in a superconductor out of equilibrium during a pump-probe experiment. For a \textit{d}-wave superconductor we find a Higgs mode at a single frequency equal to twice the maximum renormalized gap size, which is an important result for pump-probe experiments performed to study the Higgs mode in \textit{d}-wave superconductors. The parameters chosen in the simulation for the pump and probe pulses are already feasible in current trARPES setups. Under these conditions, we predict that the Higgs mode can be detected in trARPES experiments as oscillations in the spectral intensity between the energy scales set by the gap edge and the energy of the pair boson. 

\section{ACKNOWLEDGEMENTS}

We would like to thank P. Kirchmann for helpful discussions. B.N., B.M., and T.P.D. were supported by the U.S. Department of Energy, Office of Basic Energy Sciences, Materials Sciences and Engineering under Contract No. DE-AC02-76SF00515. J.K.F. was supported by the U.S. Department of Energy, Office of Basic Energy Sciences, Materials Sciences and Engineering under Contract No. DE-FG02-08ER46542 and also by the McDevitt bequest at Georgetown. Computational resources were provided by the National Energy Research Scientific Computing Center supported by the U.S. Department of Energy, Office of Science, under Contract No. DE-AC02-05CH11231.


\begin{thebibliography}{9}

\bibitem{Raman}
R. Sooryakumar and M. V. Klein, Physical Review Letters \textbf{45}, 660 (1980).

\bibitem{Raman2}
M.-A. Measson, Y. Gallais, M. Cazayous, B. Clair, P. Rodiere, L. Cairo, and A. Sacuto, Phys. Rev. B. \textbf{89}, 060503(R) (2014). 

\bibitem{Shimano}
R. Matsunaga, N. Tsuji, H. Fujita, A. Sugioka, K. Makise, Y. Uzawa, H. Terai, Z. Wang, H. Aoki, and R. Shimano, Science \textbf{345}, 1145 (2014).

\bibitem{Podolsky}
D. Podolsky, A. Auerbach, and D. P. Arovas, Phys. Rev. B \textbf{84}, 174522 (2011).

\bibitem{Varma}
D. Pekker and C. M. Varma, Annual Review of Condensed Matter Physics \textbf{6}, 269 (2015).

\bibitem{CDW}
F. Schmitt, P. S. Kirchmann, U. Bovensiepen, R. G. Moore, L. Rettig, M. Krenz, J.-H Chu, N. ru, L. Perfetti, D. H. Lu, M. Wolf, I. R. Fisher, and Z.-X. Shen, Science \textbf{321}, 1649 (2008). 

\bibitem{QQ1}
A. F. Volkov and S. M. Kogan, Zh. Eksp. Teor. Fiz. \textbf{65}, 2038 (1973) [Sov.Phys.-JETP \textbf{38}, 1018 (1974)].

\bibitem{Lex}
A. F. Kemper, M. A. Sentef, B. Moritz, J. K. Freericks, and T. P. Devereaux, Phys. Rev. B \textbf{92}, 224517 (2015).


\bibitem{ShenInequivalence}
S.-L. Yang, J. A. Sobota, D. Leuenberger, Y. He, M. Hashimoto, D. H. Lu, H. Eisaki, P. S. Kirchmann, and Z.-X. Shen, Phys. Rev. Lett. \textbf{114}, 247001 (2015).


\bibitem{Lanzara1}
J. Graf, C. Jozwiak, C. L. Smallwood, H. Eisaki, R. A. Kaindl, D.-H. Lee, and A. Lanzara, Nature Physics \textbf{7}, 805-809 (2011).

\bibitem{Lanzara2}
C. L. Smallwood, J. P. Hinton, C. Jozwiak, W. Zhang, J. D. Koralek, H. Eisadi, D.-H. Lee, J. Orenstein, and A. Lanzara, Science \textbf{336}, 1137 (2012).

\bibitem{Rameau}
J. D. Rameau, S. Freutel, M. A. Sentef, A. F. Kemper, J. K. Freericks, I. Avigo, M. Ligges, L. Rettig, Y. Yoshida, H. Eisaki, J. Schneeloch, R. D. Zhong, Z. J. Xu, G. D. Gu, P. D. Johnson, and U. Bovensiepen, ArXiv e-prints (2015), arXiv: 1505.07055 [cond-mat.supr-con].

\bibitem{Rameau2}
J. D. Rameau, S. Freutel, L. Rettig, I. Avigo, M. Ligges, Y. Yoshida, H. Eisaki, J. Schneeloch, R. D. Zhong,Z. J. Xu, G. D. Gu, P. D. Johnson, and U. Bovensiepen, Phys. Rev. B \textbf{89}, 115115 (2014).

\bibitem{Bovensiepen}
I. Avigo, S. Thirupathaiah, M. Ligges, T. Wolf, J. Fink, U. Bovensiepen, New J. Phys. \textbf{18}, 0932028 (2016).

\bibitem{Perfetti}
C. Piovera, Z. Zhang, M. d'Astuto, A. Taleb-Ibrahimi, E. Papalazarou, M. Marsi, Z. Z. Li, H. Raffy, and L. Perfetti, Phys. Rev. B \textbf{91}, 224509 (2015).

\bibitem{Shin}
Y. Ishida, T. Saitoh, T. Mochiku, T. Nakane, K. Hirata and S. Shin, Sci. Rep. \textbf{6}, 18747 (2016).

\bibitem{Cavalleri}
J. C. Petersen, S. Kaiser, N. Dean, A. Simoncig, H. Y. Liu, A. L. Cavalieri, C. Cacho, I. C. E. Turcu, E. Springate, F. Frassetto, L. Poletto, S. S. Dhesi, H. Berger, and A. Cavalleri, Phys. Rev. Lett. \textbf{107}, 177402 (2011).


\bibitem{Beyer}
M. Beyer, D. Stadter, M. Beck, H. Schafer, V.V. Kabanov, G. Logvenov, I. Bozovic, G. Koren, and J. Demsar, Phys. Rev. B. \textbf{83}, 214515 (2011).

\bibitem{Madan}
I. Madan, T. Kurosawa, Y. Toda, M. Oda, T. Mertelj, P. Kusar, D. Mihailovic, Sci. Rep. \textbf{4}, 5656 (2014).


\bibitem{Torchinsky}
D.H. Torchinsky, G.F. Chen, J.L. Luo, N.L. Wang, and N. Gedik, Phys. Rev. Lett. 105, 027005 (2010).

\bibitem{Chia}
E.E.M. Chia, D. Talbayev, J.X. Zhu, J.D. H. Thompson, A.J. Taylor, H.Q. Yuan, T. Park, C. Panagopoulos, G.F. Chen, J.L. Luo, and N.L. Wang, Phys. Rev. Lett. 104, 027003 (2010).


\bibitem{Mansart}
B. Mansart, D. Boschetto, A. Savoia, F. Rullier-Albenque, F. Bouquet, E. Papalazarou, A. Forget, D. Colson, A. Rousse, and M. Marsi, Phys. Rev. B. \textbf{82}, 024513 (2010).

\bibitem{QQ2}
E. A. Yuzbashyan, O. Tsyplyatyev, and B. L. Altshuler, Phys. Rev. Lett. \textbf{96}, 097005 (2006).

\bibitem{Aoki}
N. Tsuji and Hideo Aoki, Phys. Rev. B \textbf{92}, 064508 (2015).

\bibitem{Capone}
F. Peronaci, M. Schiro, and M. Capone, Phys. Rev. Lett. \textbf{115}, 257001 (2015).

\bibitem{VarmaDwave}
Y. Barlas and C. M. Varma, Phys. Rev. B \textbf{87}, 054503 (2013).

\bibitem{Holstein}
T. Holstein, Ann. Phys. \textbf{8}, 325 (1959).

\bibitem{Freericks}
J. K. Freericks, H. R. Krishnamurthy, and T. Pruschke, Phys. Rev. Lett \textbf{102}, 136401 (2009). 

\bibitem{Bertoncini}
R. Bertoncini and A. P. Jauho, Phys. Rev. B \textbf{44}, 3655 (1991).

\bibitem{supplement}
See Supplemental Material at \url{http://link.aps.org/supplemental/10.1103/PhysRevB.96.184518} for details of the approach, as well as plots of the self-energy in and out of equilibrium. 

\bibitem{LexPRB90}
A. F. Kemper, M. A. Sentef, B. Moritz, J. K. Freericks, and T. P. Devereaux, Phys. Rev. B \textbf{90}, 075126 (2014).

\bibitem{Leeuwen}
G. Stefanucci and R. van Leeuwen, \textit{Nonequilibrium Many-body Theory of Quantum Systems: A Modern Introduction} (Cambridge University Press, New York, 2013).

\bibitem{Sandvik}
A. W. Sandvik, D. J. Scalapino, and N. E. Bickers, Phys. Rev. B \textbf{69}, 094523 (2004).

\bibitem{DevereauxReview}
T. Cuk, D. H. Lu, X. J. Zhou, Z.-X. Shen, T. P. Devereaux, and N. Nagaosa, Phys. Stat. Sol. (b) \textbf{242}, 11 (2005).

\bibitem{Benfatto}
T. Cea and L. Benfatto, Phys. Rev. B \textbf{90}, 224515 (2014).

\bibitem{ShuolongPhonon}
S.-L. Yang, J. A. Sobota, D. Leuenberger, A. F. Kemper, J. J. Lee, F. T. Schmitt, W. Li, R. G. Moore, P. S. Kirchmann, and Z.-X. Shen, Nano Lett. \textbf{15}, 4150 (2015).

\bibitem{LanzaraGS}
C.L. Smallwood, W. Zhang, T.L. Miller, G. Affeldt, K. Kurashima, C. Jozwiak, T. Noji, Y. Koike, H. Eisaki, D.-H. Lee, R.A. Kaindl, and A. Lanzara, Phys. Rev. B \textbf{92}, 161102(R) (2015).

\bibitem{Inna}
I. M. Vishik, M Hashimoto, R.-H. He, W. S. Lee, F. Schmitt, D. H. Lu, R. G. Moore, C. Zhang, W. Meevasana, T. Sasagawa, S. Uchida, K. Fujita, S. Ishida, M. Ishikado, Y. Yoshida, H. Eisaki, Z. Hussain, T. P. Devereaux, and Z.-X. Shen, PNAS \textbf{109}, 18332 (2012).


\bibitem{Wolf}
C. Monney, M. Puppin, C.W. Nicholson, M. Hoesch, R.T. Chapman, E. Springate, H. Berger, A. Magrez, C. Cacho, R. Ernstorfer, and M. Wolf, Phys. Rev. B \textbf{94}, 165165 (2016). 

\bibitem{DevereauxBS}
T. Bohm, A. F. Kemper, B. Moritz, F. Kretzschmar, B. Muschler, H.-M. Eiter, R. Hackl, T. P. Devereaux, D. J. Scalapino, and H.-H. Wen, Phys. Rev. X \textbf{4}, 041046 (2014).

\bibitem{LeggettSchnyder}
H. Krull, N. Bittner, G. S. Uhrig, D. Manske, and A. P. Schnyder, Nature Communications \textbf{7}, 11921 (2016).

\bibitem{LeggettHuang}
W. Huang, T. Scaffidi, M. Sigrist, and C. Kallin, ArXiv e-prints (2016), arXiv: 1605.03800v2 [cond-mat.supr-con].   

\bibitem{Lee}
P. A. Lee, N. Nagaosa, Phys. Rev. B \textbf{68}, 024516 (2003).

\end{thebibliography}
\end{document}